\documentclass[12pt]{article}
\usepackage{graphicx}
\usepackage{epsfig}
\usepackage{float}
\begin{document}
\rightline{NKU-2017-SF2}
\bigskip

%%%%%%%%%%%new commands%%%%%%%%%%%%%%%%%

\newcommand{\be}{\begin{equation}}
\newcommand{\ee}{\end{equation}}
\newcommand{\noi}{\noindent}
\newcommand{\refb}[1]{(\ref{#1})}
\newcommand{\ra}{\rightarrow}
\newcommand{\bib}{\bibitem}
\newcommand{\bigd}{\bigtriangledown}

%%%%%%%%%%%%%%%%%%%Title%%%%%%%%%%%%%%%%%%
\begin{center}
{\Large\bf Quasinormal modes and absorption cross sections  of Born-Infeld-de Sitter black holes}
\end{center}
\hspace{0.4cm}

%%%%%%%%%%%%%%%%%Authors%%%%%%%%%%%%%%%%
\begin{center}
Nora Bret\'on \footnote{nora@fis.cinvestav.mx},\\
{\small \it Dpto de F\'isica, Centro de Investigaci\'on y de Estudios Avanzados \\
del I.P.N., Apdo. 14-740, Mexico City, Mexico.}\\
\vspace{0.3 cm}
Tyler Clark \footnote{clarkt3@mymail.nku.edu}, \& Sharmanthie Fernando \footnote{fernando@nku.edu}\\
{\small\it Department of Physics, Geology  \& Engineering Technology}\\
{\small\it Northern Kentucky University}\\
{\small\it Highland Heights, Kentucky 41099, U.S.A.}\\
\end{center}

%%%%%%%%%%%Abstract%%%%%%%%%%%%%%

\begin{abstract}

In this paper, we have studied  QNM modes and  absorption cross sections of  Born-Infeld-de Sitter black holes. WKB approximation is employed to compute the QNM modes of 
 massless scalar fields. We have also used  null geodesics to compute  quasi-normal modes in the eikonal approximation. In the eikonal limit QNMs of black holes are
determined by the parameters of the circular null geodesics.
Unstable circular null orbits are derived from the effective metric
which is  obeyed by  light rays under the influence of a
nonlinear electromagnetic field. Comparison is shown with the QNM of the 
linear electromagnetic counterpart, the Reissner-Nordstr\"{o}m black hole. Furthermore, the null geodesics are employed to compute the absorption cross sections in the high frequency limit via the sinc approximation.

\end{abstract}


%%%%%%%%%%%%%%%%Introduction%%%%%%%%%%%%%%

\section{Introduction}
It is well known that the self-energy  and the electric field diverges for a point charge in Maxwell's electrodynamics. In 1934, Born and Infeld developed a theory of non-linear electrodynamics in order to obtain a finite value for the self-energy of a point charge \cite{born}. The action for the Born-Infeld electrodynamics coupled to gravity with a cosmological constant is given by,

\begin{equation} \label{action}
S = \int d^4x \sqrt{-g} \left[ \frac{R - 2 \Lambda)}{16 \pi G} + L(F) \right]
\end{equation}
where the  function  $L(F)$ is given by,
\begin{equation}
L(F) = 4 b^2 \left( 1 - \sqrt{ 1 + \frac{F}{ 2 b^2} - \frac{ G^2}{ 16 b^4}} \right)
\end{equation}
Here, $ F =   F^{\mu \nu}F_{\mu \nu}$ and $ G =  \tilde{F}^{\mu \nu}F_{\mu \nu}$. The dual tensor $\tilde{F}_{\mu \nu}$ is given by,
\begin{equation}
\tilde{F}_{\mu \nu} = \frac{\epsilon_{\mu \nu \rho \sigma} F^{ \rho \sigma}}{ 2 \sqrt{-g}}
\end{equation}
$b$ in the action is a parameter of the Born-Infeld non-linear electrodynamics and has the dimensions $length^{-2}$. 
When $b \rightarrow \infty$, the Lagrangian $L(F) \rightarrow -F$ which corresponds to  Maxwell's  electrodynamics. In flat space, the electric field strength $E$ in Born-Infeld electrodynamics is given by, $E = Q/ \sqrt{ r^4 + \frac{Q^2}{b^2}}$ which is finite at $r =0$. Among the appealing features of Born-Infeld electrodynamics include, the finiteness of energy density of charged particles, its relation to string theory, the electro magnetic duality \cite{rasheed1}, its exceptional properties of wave propagation and the absence of birefringes \cite{guy} \cite{yuri}.

In recent years,  Born-Infeld theory has attracted considerable attention due to its role in string theory. For example, Leigh \cite{leigh} showed that the low energy dynamics of D-branes could be described by a Dirac-Born-Infeld type  action.  The action for a massless vector field in open superstring theory is given by a modified Born-Infeld action as discussed by Tseytlin et.al.  \cite{tsey1} \cite{tsey2}. Bergshoeff et.al. \cite{berg} showed that one-loop approximation to sigma-model perturbation theory of gauge fields in  open-superstrings lead to the Born-Infeld action. A review of the Born-Infeld theory and string theory can be found in \cite{gib1}.

There are many interesting black hole solutions and physics related  to Born-Infeld electrodynamics in the literature. Asymptotically flat charged black hole solutions in Born-Infeld gravity and the geodesics  were presented by Bret$\acute{o}$n in \cite{nora1}. Test particle motion around Born-Infeld black holes were studied by Linares et.al. \cite{lin}. Null geodesics of Born-Infeld black holes were studied by Fernando \cite{fer7}.

Stability and quasi-normal modes  of the asymptotically flat charged black hole in Born-Infeld gravity were studied by Fernando et.al. in \cite{fer1}\cite{fer2}\cite{fer3}. Nariai black holes of Born-Infeld black holes in de Sitter space were studied by Fernando \cite{fer4}. Energy decomposition in Born-Infeld charged black hols were presented in \cite{pere} and the energy extraction from the Born-Infeld black holes was studied in \cite{nora2}

Melvin universe in  Born-Infeld electrodynamics was studied by Gibbons and Herderio \cite{gib2}. Charged black hole solutions in Lovelock-Born-Infeld black holes in 5 dimensions were studied by Aiello et.al \cite{ello}.  Attractor mechanism for extreme black hole solutions in Einstein-Born-Infeld-dilaton gravity was studied in  \cite{gao}.  There are several works to present black hole solutions in Born-Infeld theory coupled to Einstein gravity with a cosmological constant  \cite{cat}\cite{fer5}\cite{dey}\cite{cai}. 

%%%%%%%%QNM%%%%%%%%%%%

Studying quasi normal modes (QNM) of a black hole is an well established subject. When a black hole is perturbed by fields of various spins or by perturbing the metric, there are oscillations with complex frequencies. Such frequencies depend only on the black hole and the  corresponding field parameters. When two black holes merge,  as the historic event that occurred in 2015 creating gravitational waves, the intermediate signals correspond to the QNM's \cite{ligo}. Hence, it would be important to calculate QNM frequencies in alternative theories such as general relativity coupled to Born-Infeld theory to see if there are deviations from the usual general relativity coupled to Maxwell's theory.
Hence there is  strong motivation to study and calculate QNM's of black holes in theories such as Born-Infeld-de Sitter black holes. There are many works in the literature to compute QNM frequencies of variety of perturbations. For example, QNM frequencies of canonical acoustic holes (black hole analogue)  were calculated by Dolan et.al.\cite{dolan}. QNM's of regular black hole with a magnetic charge were calculated by Li et.al.\cite{li}.  A comparison of different cosmological models were done using QNM's of the Schwarzschild-de Sitter black hole by  Chirenti et.al.\cite{chi}. QNM's of Schwarzschild black holes were calculated by a matrix method by Lin and Qian\cite{lin3}. A ``generalized continued fraction method '' was employed to compute QNM frequencies of an extreme Reissner-Nordstrom and Kerr black hole by Richartz \cite{rich}.  A new approach, ``Asymptotic Iteration Method'' was used to calculate black hole QNM's by Cho et.al.\cite{cho}.  There are several works on QNM's of de Sitter black holes by Konoplya  \cite{Konoplya:2008au}\cite{Konoplya:2013sba}\cite{Konoplya:2007zx}\cite{Konoplya:2004uk}. There are so many works on this topic that there is not sufficient space to mention all of them here; there is an excellent review on QNM's of black holes and black branes by Berti et.al.\cite{berti}.

In this paper we are interested in  computing QNM's and absorption cross sections of Born-Infeld black hole with a cosmological constant. As is well known, the current universe is expanding at an accelerated rate \cite{perl}\cite{riess}\cite{sper}\cite{teg}\cite{sel}. The simplest explanation for the acceleration is the existence of a cosmological constant. Hence it becomes timely to study black holes  with a cosmological constant.

The paper is organized as follows: In section 2, an introduction to the Born-Infeld-de Sitter black hole is given. In section 3 the massless scalar field perturbation is introduced. In section 4, the QNM frequencies are calculated via the WKB approximation for various parameters of the theory. In section 5, the QNM's are calculated via the null geodesics of the Born-Infeld-de Sitter black hole. In section 6, the absorption cross section via the null geodesics are computed and presented. Finally, the conclusion is given in section 7.

%%%%%%%%%%%%%%%%%%%%%%%%%%%%%%%%%%%%%%%%%%%%%%%%%%%%%%
\section{Introduction to Born-Infeld-de Sitter black holes}

\begin{figure} [H]
\begin{center}
\includegraphics{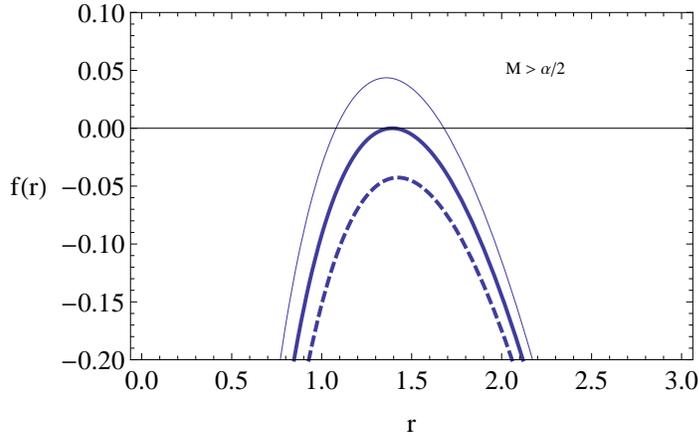}
\caption{The figure shows  $f(r)$ vs $r$ when  $ M > \frac{\alpha}{2}$. Here, $\Lambda = 0.5, b= 1.4$ and $Q = 0.2308$}
\label{schfr}
 \end{center}
\end{figure}

In this section, we will present the Born-Infeld-de Sitter black hole solutions and some important properties. The metric of the static charged, spherical symmetric black hole derived from the action in eq.$\refb{action}$ for $F \neq 0$ and $G =0$ is given by,
\begin{equation} \label{metric}
ds^2 = -f(r) dt^2 + f(r)^{-1} dr^2 + r^2 ( d \theta^2 + sin^2\theta d \varphi^2)
\end{equation}
with,
\begin{equation}
f(r)=1- \frac{2M}{r}- \frac{\Lambda r^2}{3}+  \frac{2 b^2}{3}[r^2- \sqrt{r^4+Q^2/b^2}]+ \frac{2Q^2}{3r} \sqrt{\frac{b}{Q}} F \left[{\arccos \left({\frac{r^2-Q/b}{r^2+Q/b}} \right), \frac{1}{\sqrt{2}}} \right],
\end{equation}
We have used the elliptic function $F$ in the above equation. In the literature, the function $f(r)$ is also written in terms of the Hypergeometric function  $_2F_1$. For the sake of completeness we want to point out that the relationship between the elliptic function and the hypergeometric function is given as follows:
\begin{equation}
\int_{r}^{\infty} \frac{ds}{\sqrt{Q^2/b^2+s^4}}=\frac{1}{r} {}_2F_{1}\left[{\frac{1}{4},\frac{1}{2};\frac{5}{4}; -\frac{Q^2}{b^2r^4}}\right]=\frac{1}{2 }\sqrt{\frac{b}{Q}} F \left[{\arccos \left({\frac{r^2-Q/b}{r^2+Q/b}} \right),\frac{1}{\sqrt{2}}}\right],
\label{elliptic_hypergeom}
\end{equation}
The parameters in the metric are as follows; $M$ is the mass, $Q$ is the electric charge, $b$ is the non-linear parameter and $\Lambda$ is the cosmological constant. The electric field strength for the black hole space-time is  given by,
\be
E = \frac{ Q}{ \sqrt{ r^4 + \frac{Q^2}{b^2} }}
\ee
 \noi
In the limit $b \rightarrow \infty$, the metric functin $f(r)$ can be expanded to give,
\begin{equation} \label{frn}
f(r)_{RN} = 1 - \frac{2 M}{r} + \frac{ Q^2}{r^2} - \frac{ \Lambda r^2}{3}
\end{equation}
which is the function  $f(r)$ for the Reissner-Nordstrom-de Sitter  black hole for Maxwell's electrodynamics.
When $ r \ra 0$, the function $f(r)$ has the behavior,
\begin{equation}
f(r) \approx 1 - \frac{( 2M - \alpha)}{r} - 2 b Q + \frac{ r^2}{3}  ( 2 b^2 - \Lambda)  +  \mathcal{O} (r^3)
\end{equation}
Where,
\begin{equation}
\alpha = \frac{1}{3}\sqrt{ \frac{b}{ \pi} } Q^{3/2} \Gamma \left(\frac{1}{4} \right)^2
\end{equation}
\noi
The singular nature of the black holes depends on  the relative values of the parameters $M$ and $\alpha$, as described below: \\
\noi
{\bf Case 1}: ($M > \frac{\alpha}{2}$)\\
\noi
In this case, for $ r \ra 0$, $ f(r) \ra - \infty$;   the black hole will behave  similar to the Schwarzschild-de Sitter black hole as given in the Fig$\refb{schfr}$. It is possible to have two horizons, an event horizon ($r_h$) and a cosmological horizon($r_c$). It is also possible to have extreme black holes with one degenerate horizon. For large masses, there won't be any horizons and there will be naked singularities.\\

\begin{figure} [H]
\begin{center}
\includegraphics{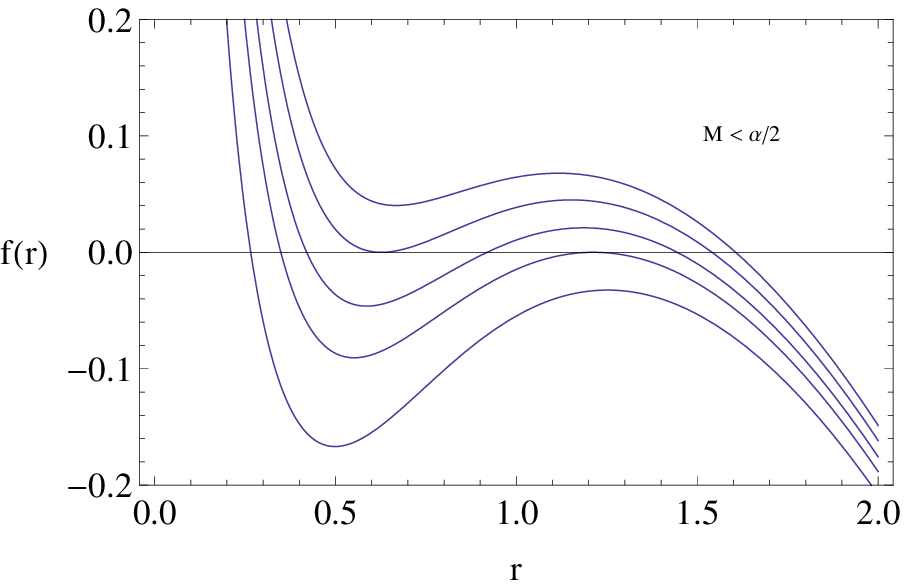}
\caption{The figure shows  $f(r)$ vs $r$ when  $ M < \frac{\alpha}{2}$. Here, $\Lambda = 0.5, b= 1.4$ and $Q = 0.63079$}
\label{rnfr}
 \end{center}
\end{figure}

\noi
{\bf Case 2}: ($ M < \frac{\alpha}{2}$)\\
\noi
Here, for $ r \ra 0$, $ f(r) \ra  \infty$; the behavior of  the black hole will be similar to the Reissner-Nordstrom-de Sitter black hole (RNdS black hole) as shown in Fig$\refb{rnfr}$. As one can observe, there are many possibilities for this geometry. First, it is possible to have three horizons with the largest one being the cosmological constant. It is possible to have extreme black holes where there are only two horizons exists. For special parameters, a degenerate horizon exists with one horizon. All these possibilities are given in Fig.$\refb{rnfr}$.\\

\begin{figure} [H]
\begin{center}
\includegraphics{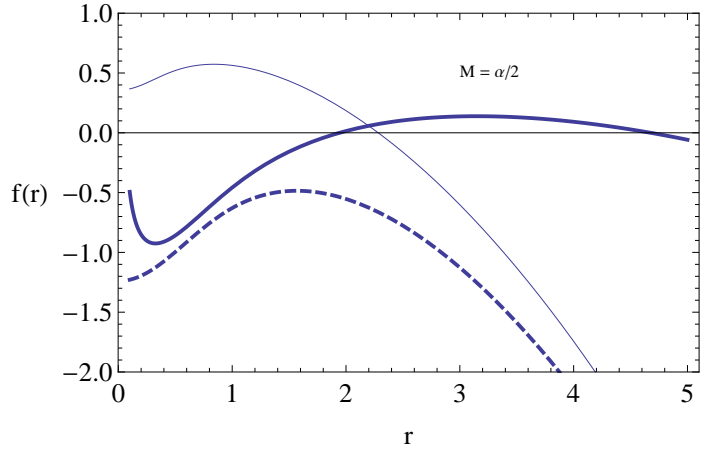}
\caption{The figure shows  $f(r)$ vs $r$ when  $ M = \frac{\alpha}{2}$. Here, $b= 1.5$}
\label{single}
 \end{center}
\end{figure}

\noi
{\bf Case 3}: ($ M = \frac{\alpha}{2}$)\\
\noi
Here, for $ r \ra 0$, $ f(r) \ra  ( 1- 2 Q \beta)$; $f(r)$ is finite at $ r =0$ and single valued as given in Fig.$\refb{single}$. The Kretschmann scalar  diverge at $ r =0$ for these black holes; so, the singularity exits. 

\begin{figure} [H]
\begin{center}
\includegraphics{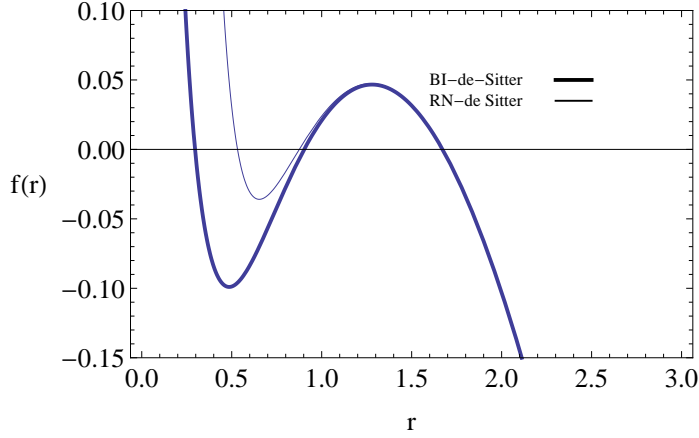}
\caption{The figure shows  $f(r)$ vs $r$ for BIdS and RNdS black holes.  Here $ M = 0.593, Q =0.6, \Lambda =0.45$ and $b= 1.4$}
\label{rnbi}
 \end{center}
\end{figure}

In comparison with the Reissner-Nordstrom-de Sitter black hole, the geometry seems to be different mostly closer to the horizons as shown in Fig$\refb{rnbi}$.

%%%%%%%%%%%%%%%equations of perturbations for the scalar field%%%%%%%%%%%%%%%%%%%%%

\section{ Massless scalar field perturbations}

In this section, we will focus on  perturbations by a minimally coupled massless scalar field   in the fixed background of the Born-Infeld-de Sitter black hole (BIdS black hole). The equations of motion of the minimally coupled scalar field is given by the Klein-Gordon equation, $\bigtriangledown^2 \Phi = 0$. The scalar filed $\Phi$ can be decomposed to its partial modes  in terms of spherical harmonics, $Y_{l,m}(\theta, \phi)$ as,
\be \label{decom}
\Phi(r, \theta, \phi,t) = \sum_{l,m}  e^{ - i \omega t}  Y_{ l, m} ( \theta, \phi) \frac{ R(r)}{ r}
\ee
Here,  $l$ and $m$ are the angular and the magnetic quantum numbers respectively. $\omega$ is the oscillating frequency of the scalar field. With the decomposition in eq$\refb{decom}$, the  Klein-Gordon equation will separate into radial equation given by,
\be \label{wave}
\frac{ d^2 R(r_*) }{ dr_*^2} + \left( \omega^2 - V_{scalar}(r_*)  \right) R(r_*) = 0
\ee
Here  $r_*$ is the tortoise coordinates which is given by, $dr_* = \frac{ dr} { f(r)}$ and  $V_{scalar}$ is given by,
\be \label{potscalar}
V_{scalar}(r) = f(r) \left( \frac{ l ( l + 1) } { r^2} + \frac{   f'(r)}{r}  \right)
\ee

%%%%%%%%%%%%en perturbations%%%%%%%%%%%%
The function $f(r)$ is zero at the event horizon, $r_h$ and at the cosmological horizon $r_c$. Therefore, $V_{scalar}(r_h) = V_{scalar}(r_c) =0$. It is not possible to calculate $R(r_*)$ analytically; we can get an approximation by expanding $f(r)$ around $r=r_h$ and $r= r_c$  in a Taylor series as,
\be
f(r)  \approx f'(r_{h}) ( r - r_{h})
\ee
\be
f(r)  \approx f'(r_{c}) ( r - r_{c})
\ee
Since $ f'( r_c) < 0$, $r_*$ closer to $r= r_c$ can be written as,
\be \label{torcos}
r_* \approx   - \frac{1}{ |f'(r_c)|} Log( r_c - r)
\ee
Hence, for $r \ra r_c$, $r_* \ra \infty$.
Since  $ f'(r_h) >0$,  $r_*$  closer to $r=r_h$ can be written as,
\be
r_* \approx \frac{1}{ |f'(r_h)|} Log( r - r_h)
\ee
Hence, for $r \ra r_h$, $r_* \ra - \infty$.
Hence the effective potential $V_{scalar} \ra 0$ when $ r_* \ra \pm \infty$.

%%%%%%%%%%%%%%%%%%%WKB approximation%%%%%%%%%%%%%%%%%%%%%%

\section{ QNM frequencies  of the scalar field by the WKB approximations}

QNM frequencies of the scalar perturbations can be computed by imposing proper boundary conditions on the solutions of the eq$\refb{wave}$. Usually, eq$\refb{wave}$ cannot be solved analytically. There are few exactly solvable perturbation equations in the literature. In 2 +1 dimensions, exact QNM values have been found  for black holes with a dilaton field by Fernando \cite{fernando2}\cite{fernando3}\cite{fernando4}.

The boundary conditions imposed for an asymptotically de Sitter black hole space-time  goes like this: the field has to be purely ingoing at the black hole event horizon, $r_h$ and the field has to be purely outgoing at the cosmological horizon, $r_c$. With such boundary conditions, $R(r_*)$  behaves as,
\be
R(r_*) \ra exp( i \omega r_*);  \hspace{1 cm}  r_* \ra - \infty ( r \ra r_h)
\ee
\be
R(r_*) \ra exp( -i \omega r_*);  \hspace{1 cm}  r_* \ra + \infty ( r \ra r_c)
\ee
When such boundary conditions are imposed, the resulting frequencies, $\omega$ are complex. As it is clear from section(4.1), the effective potential $V_{scalar}$ has a peak and acts as a barrier to the scalar field in the background of the black hole geometry. In this scenario one can employ the WKB approximation to compute $\omega$. Iyer and Will developed it for third order in \cite{will} and was extended to sixth order by Konoplya in \cite{kono4}. WKB approximation has been employed to study QNM in many papers including  \cite{fernando5}  \cite{fernando6} \cite{gauss} .

In the sixth order WKB approach,   QNM frequencies are given by the expression,
\be
\omega^2 = - i \sqrt{ - 2 V''(r_{m})} \left( \sum^6_{i=2}  \mathcal{T}_i  +   n + \frac{ 1}{2} \right) + V(r_{m})
\ee
Here $V(r)$ is the effective potential for the corresponding equations, $r_{m}$ is  where $V(r)$ reach a  maximum and $V''(r)$ is the second derivative of the potential.  Expressions for $\mathcal{T}_i$ can be found in \cite{kono4}. 

%%%%%%%%%%%%%%%%%%%%%
\subsection{ Effective potentials}

\begin{figure} [H]
\begin{center}
\includegraphics{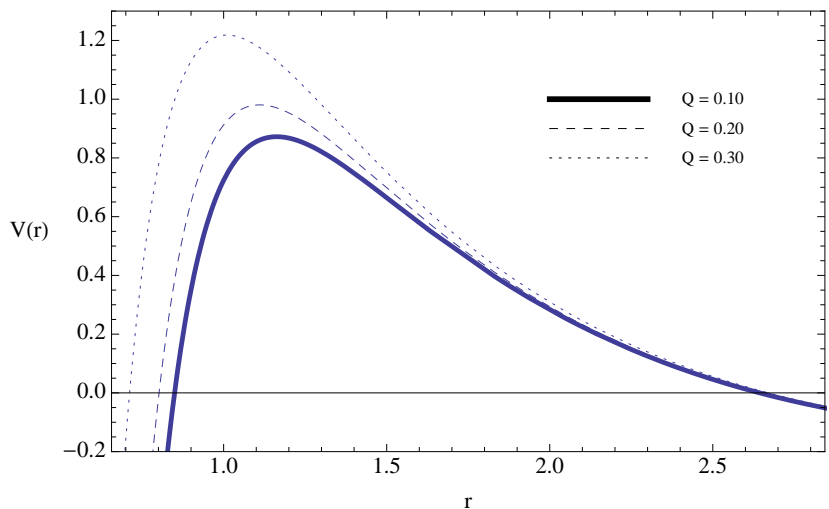}
\caption{The figure shows  $V(r)$ vs $r$ for massless scalar field for varying charge $Q$. Here, $\Lambda = 0.3, b= 1, M = 0.4, l=2$}
\label{potq}
 \end{center}
\end{figure}

\begin{figure} [H]
\begin{center}
\includegraphics{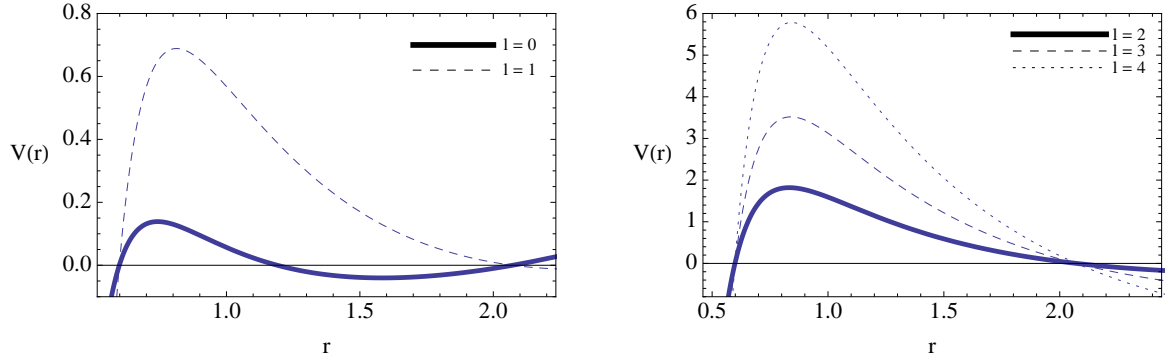}
\caption{The figure shows  $V(r)$ vs $r$ for massless scalar field for varying $l$. Here, $\Lambda = 0.5, b= 1, M = 0.3, Q = 0.15$}
\label{potl}
 \end{center}
\end{figure}

\begin{figure} [H]
\begin{center}
\includegraphics{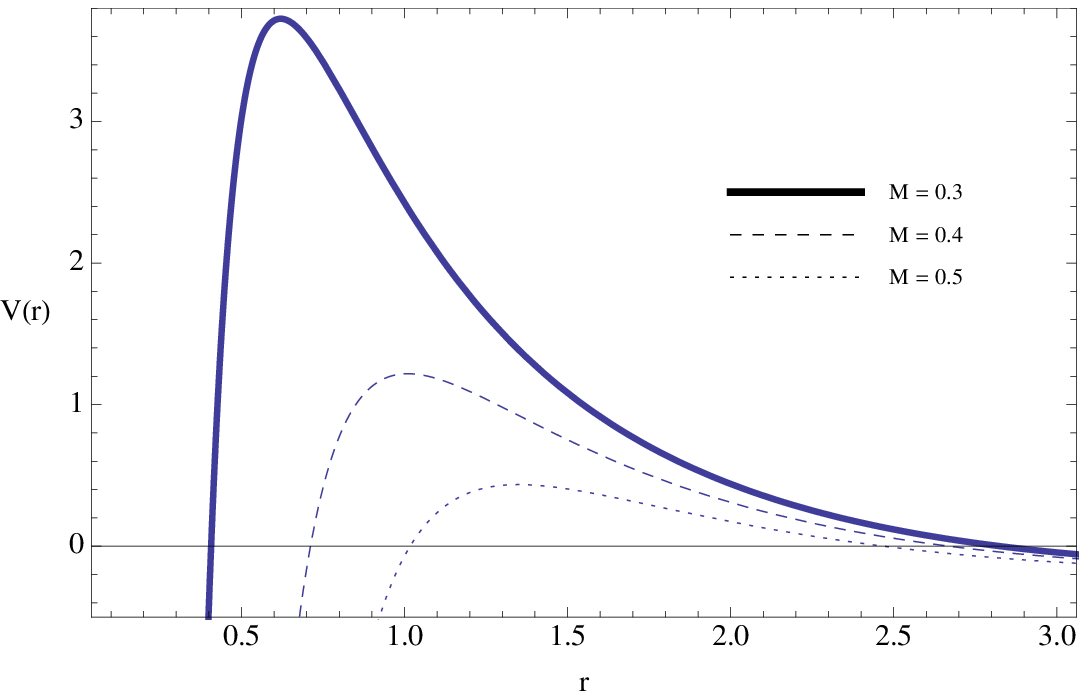}
\caption{The figure shows  $V(r)$ vs $r$ for massless scalar field for varying $M$. Here, $\Lambda = 0.3, b= 1, Q = 0.3, l=2$}
\label{potmass}
 \end{center}
\end{figure}

\begin{figure} [H]
\begin{center}
\includegraphics{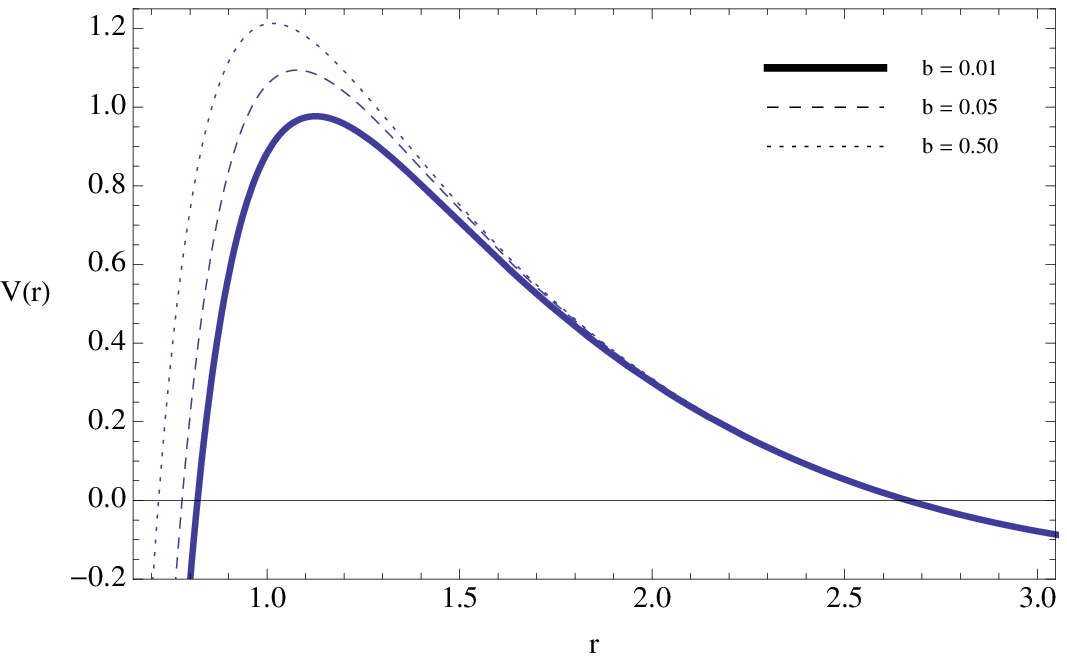}
\caption{The figure shows  $V(r)$ vs $r$ for massless scalar field for varying $b$. Here, $\Lambda = 0.3, Q = 0.3, M =0.4, l=2$}
\label{potb}
 \end{center}
\end{figure}

\begin{figure} [H]
\begin{center}
\includegraphics{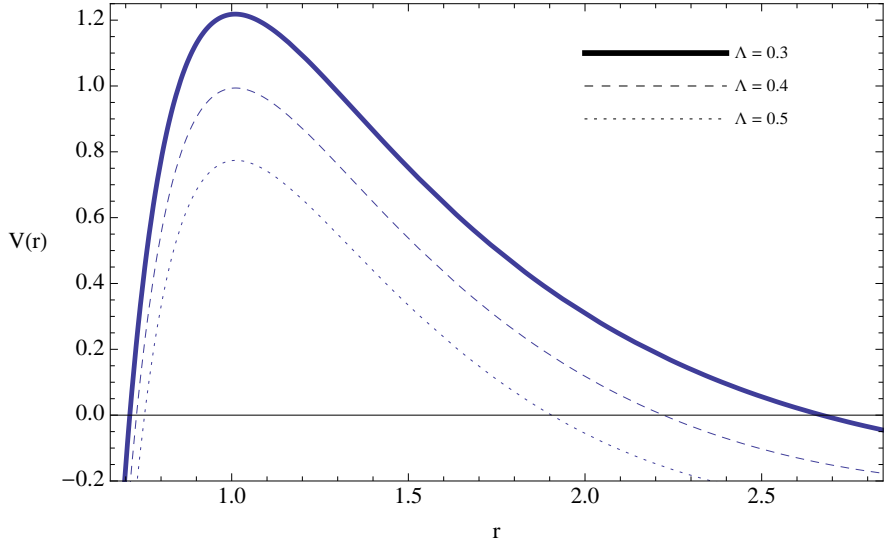}
\caption{The figure shows  $V(r)$ vs $r$ for massless scalar field for varying $\Lambda$. Here, $Q = 0.3, M =0.4,l=2, b=1$}
\label{potlambda}
 \end{center}
\end{figure}

In this section we will focus on the properties of the effective potential of the scalar field.  $V_{scalar}$ is plotted for the parameters in the theory, $M, Q, l, b$ and $\Lambda$. In Fig$\refb{potq}$, the potential is plotted for various values of the charge $Q$. When $Q$ increases, the height of the potential increases. In Fig$\refb{potl}$, the potential is plotted for various values of $l$.  For $l >0$, the potential is positive in the region between the horizons and the  height of the potential increases with $l$. On the left hand side, the potential is plotted for $l=0$. It has a local minimum between the horizons and is negative in one part of the region. Such behavior for $l=0$ scalar field potential is common for black holes with a positive cosmological constant \cite{fernando5}. In Fig$\refb{potmass}$, the potential is plotted by varying the mass. When the mass is higher, the height of the potential is lower. In Fig$\refb{potb}$, the potential is plotted by varying the non-linear parameter $b$. When $b$ increases, the  height increases. Hence one can conclude that the height of the potential of the  Reissner-Nordstrom-de Sitter black hole is higher than the Born-Infeld-de Sitter black hole with the same charge. In Fig$\refb{potlambda}$, the potential is plotted to understand how it behaves with $\Lambda$. For higher $\Lambda$, the potential has a lower height.

%%%%%%%%%%%%%%%%%%%omega%%%%%%%%%%

\subsection{ QNM frequencies for each parameter in the theory}

In this section, we will discuss the QNM frequencies for each parameter in the theory. It is noted that all calculated values, $\omega_i$ is negative. Hence the black hole is stable for the massless scalar field for $l >0$. Due to the nature of the potential for $l =0$ case, the WKB approximation cannot be used to calculate the QNM values for that particular case. Hence all our calculations are for $ l >0$ and the one for $ l=0$ case is left for future studies. Also, unless mentioned the 
$\omega$ is calculated for the fundamental harmonic $n=0$ case. 

\begin{figure} [H]
\begin{center}
\includegraphics{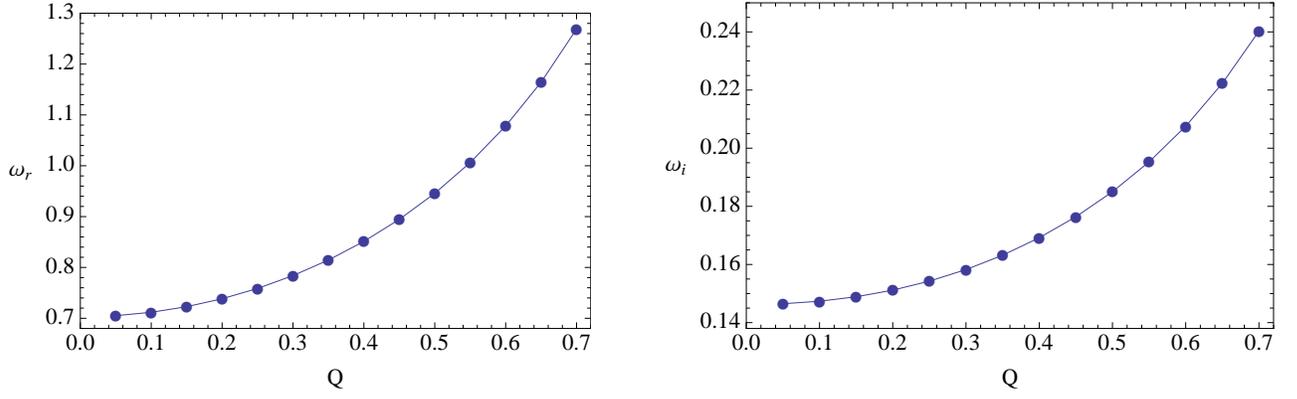}
\caption{The figure shows  $\omega$ vs $Q$ for massless scalar field. Here $M =0.5, l =2, \Lambda =0.2, \beta =0.1$}
\label{omegaq}
 \end{center}
\end{figure}

In Fig$\refb{omegaq}$, $\omega$ is presented for varying $Q$. Both $\omega_r$ and $\omega_i$ increases with $Q$. Hence the field decay faster for large charge. \\

{\bf Quality Factor}: One can define a quantity called the ``Quality Factor"  or the Q factor for any resonating system. For a standard harmonic oscillator Q factor determines the qualitative behavior of the oscillator. A system with low Q factor is said to be over damped and a system with high Q factor is said to be under damped. Along those lines, it is interesting to compare oscillatory nature of the scalar field in the black hole background in this work for various parameters. Here the Q factor is defined as \cite{frolov}
\be
Quality \hspace{0.1cm} Factor = \frac{\omega_r}{ 2 |\omega_i|}
\ee
If the Quality Factor is large, then the scalar field is  a better oscillator in the given black hole background.  

\begin{figure} [H]
\begin{center}
\includegraphics{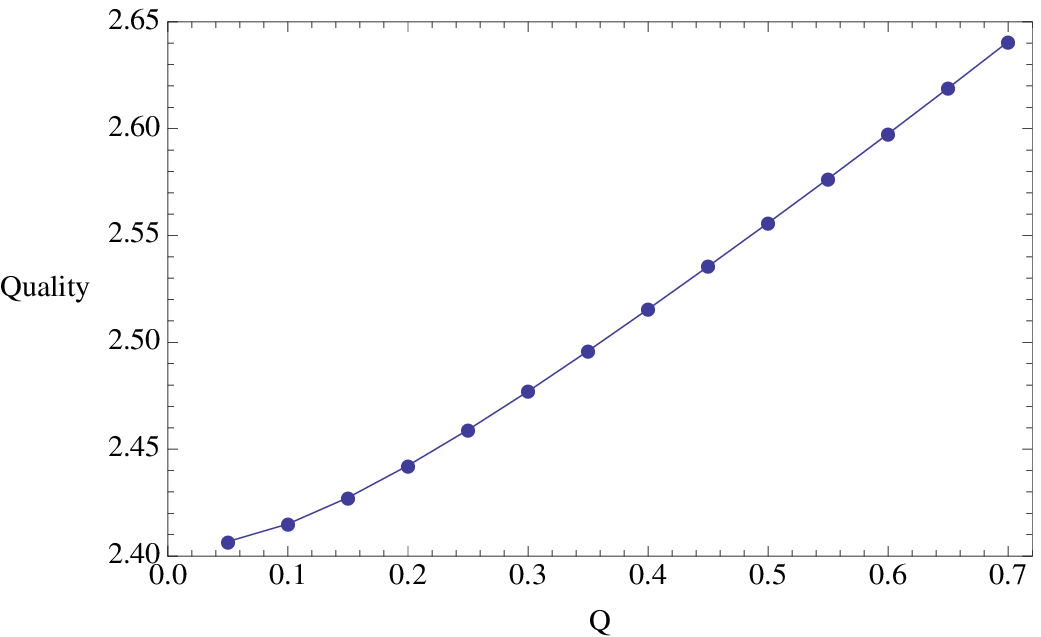}
\caption{The figure shows  Quality Factor vs $Q$ for massless scalar field}
\label{qualityq}
 \end{center}
\end{figure}

In Fig$\refb{qualityq}$, the Quality Factor is plotted for varying charge $Q$. It increases when $Q$ increases. Hence, the scalar field is a better oscillator for large charge. 

\begin{figure} [H]
\begin{center}
\includegraphics{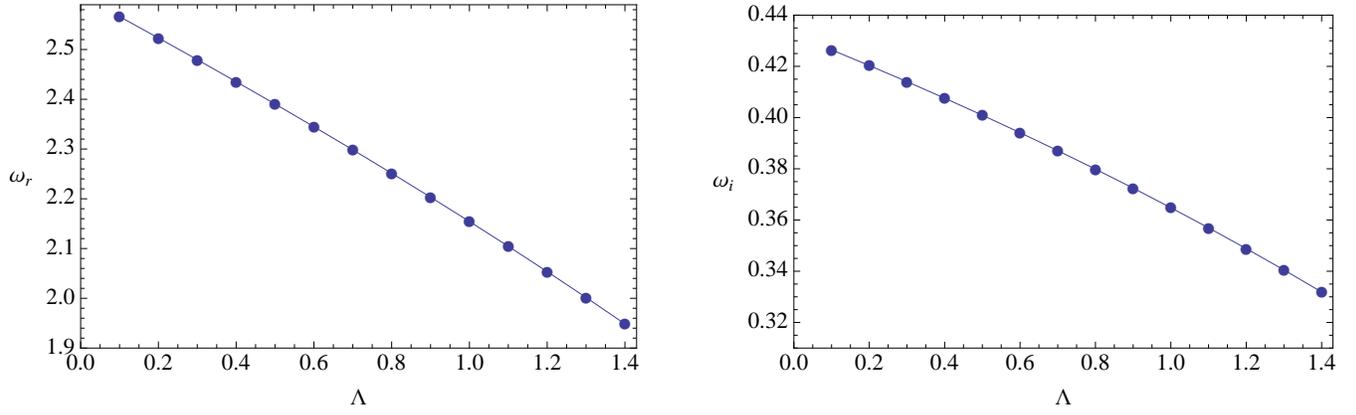}
\caption{The figure shows  $\omega$ vs $\Lambda$ for massless scalar field. Here $M =0.5, l =2, Q = 0.45, \beta =0.2$}
\label{omegalam}
 \end{center}
\end{figure}

Next we computed $\omega$ by changing the cosmological constant $\Lambda$ as is given in Fig$\refb{omegalam}$. Both $\omega_r$ and $\omega_i$ decreases as $\Lambda$ increases; a black hole with smaller $\Lambda$ is more stable for scalar perturbations. When the Quality Factor is observed, one could see that it decreases with $\Lambda$; the scalar field is a better oscillator for small $\Lambda$. 

We studied the behavior of $\omega$ with respect to $b$ in Fig$\refb{omegab}$ and Fig$\refb{qualityb}$. The real part of $\omega$ increases and reach  a stable value for large $b$. The imaginary part of $\omega$ increases to reach a maximum and then decreases to reach a stable value. Since for $b \ra \infty$, the BIdS black hole behaves like the RNdS black hole, the stable value is expected to  be the one for the RNdS value for the same parameters given by, $\omega_{RNdS} =   0.95471 - i 0.16272$. The BIdS black hole is most stable for the maximum value of $\omega_i$. Also, BIdS black hole oscillates with less frequency compared to the RNdS black hole. The Quality Factor increases with $b$. Hence the scalar field  is a better oscillator in the RNdS black hole back ground  compared to its behavior in the   BIdS black hole back ground.

Next we analyzed $\omega$ with respect to $l$; $\omega_r$ and $\omega_i$ are plotted in Fig$\refb{omegal}$ for both $ n=0$ and $ n=1$. $\omega_r$ increases linearly with $l$ where as $\omega_i$ decreases with $l$ to reach  stable value. The black hole is stable for small $l$.

In Fig$\refb{omegan}$, $\omega$ vs $n$ is plotted to establish how the black hole oscillates for higher harmonics. Since the accuracy of the WKB approximation is higher for $ l >n$, we have chosen $ l =$ for this calculation. Here $\omega_r$ decreases with $n$; the scalar field oscillates more for lower harmonics which is opposite for standing waves for a string. $\omega_i$ increases with $n$, and, hence the field decays faster for higher harmonics.

\begin{figure} [H]
\begin{center}
\includegraphics{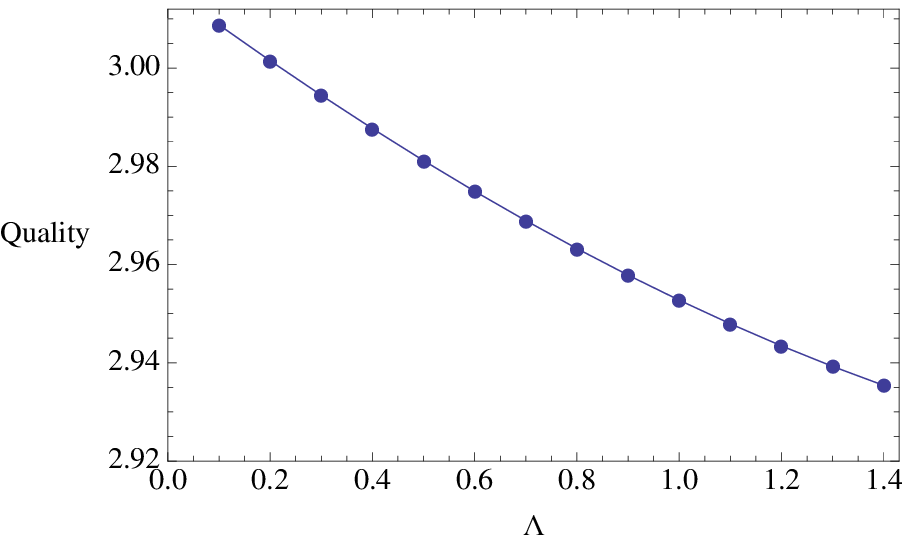}
\caption{The figure shows  Quality Factor vs $\Lambda$ for massless scalar field}
\label{qualitylam}
 \end{center}
\end{figure}

\begin{figure} [H]
\begin{center}
\includegraphics{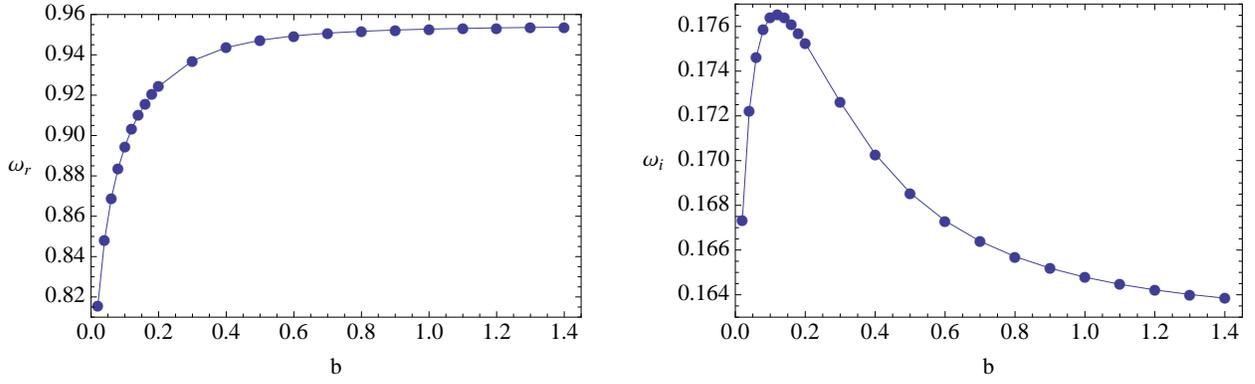}
\caption{The figure shows  $\omega$ vs $b$ for massless scalar field. Here $M =0.5, l =2, \Lambda =0.2, Q =0.45$}
\label{omegab}
 \end{center}
\end{figure}

\begin{figure} [H]
\begin{center}
\includegraphics{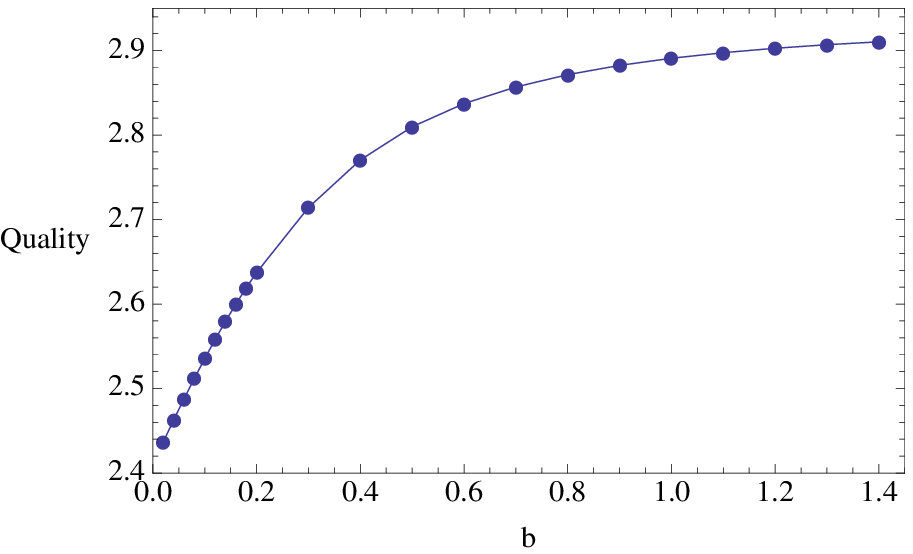}
\caption{The figure shows  Quality Factor vs $b$ for massless scalar field}
\label{qualityb}
 \end{center}
\end{figure}

\begin{figure} [H]
\begin{center}
\includegraphics{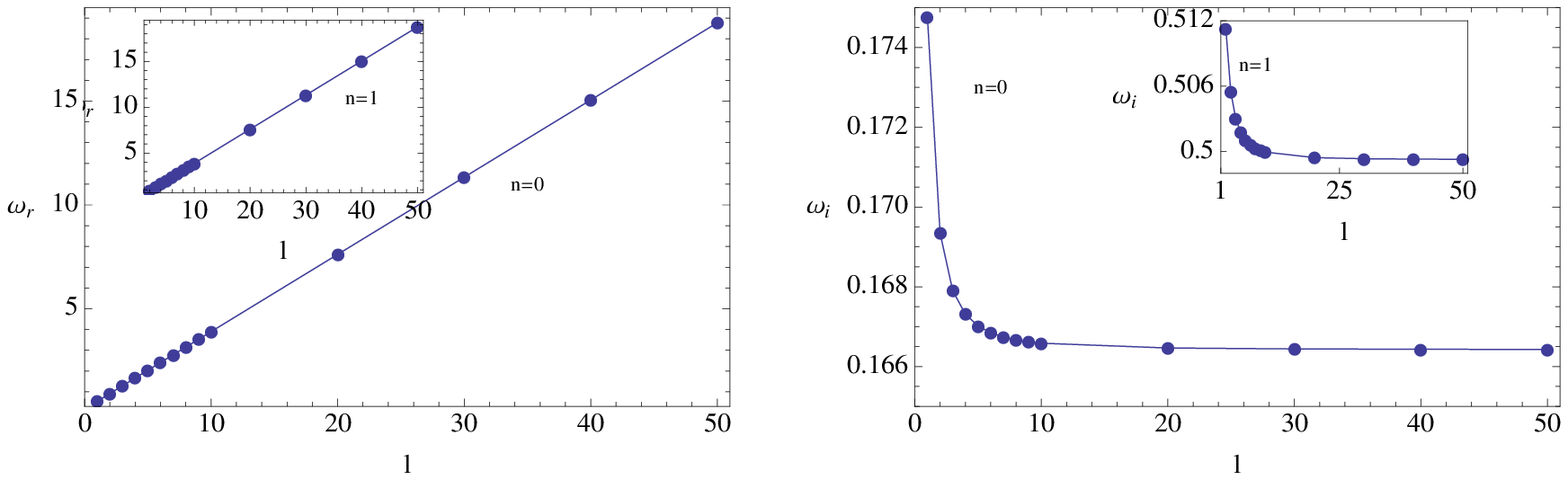}
\caption{The figure shows  $\omega$ vs $l$ for massless scalar field. Here $M =0.7, Q =1, \Lambda =0.2, \beta =0.1$}
\label{omegal}
 \end{center}
\end{figure}

\begin{figure} [H]
\begin{center}
\includegraphics{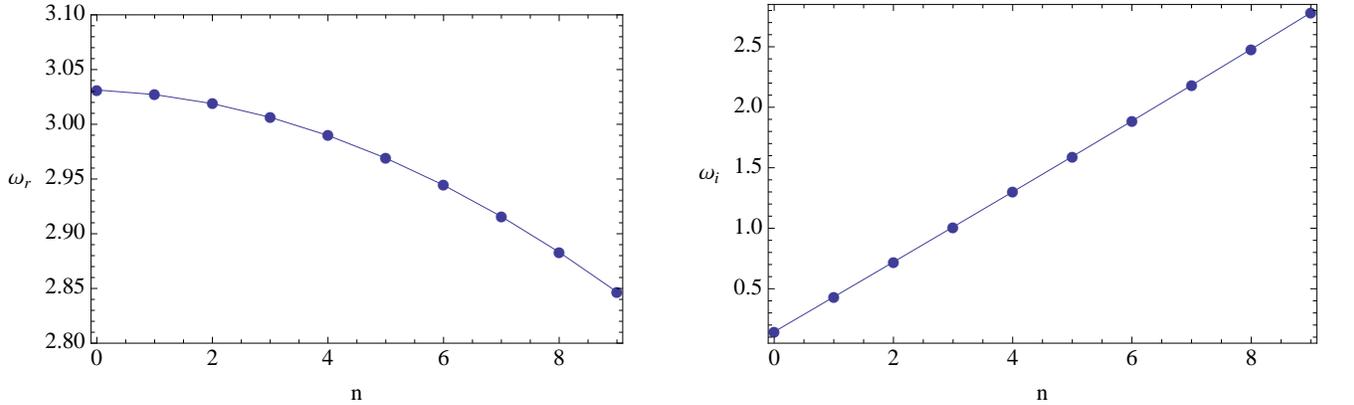}
\caption{The figure shows  $\omega$ vs $n$ for massless scalar field. Here $M =0.5, l = 10, \Lambda =0.2, \beta =0.2, Q =0.1$}
\label{omegan}
 \end{center}
\end{figure}

%%%%%%%%%%%%%%%%%%%%%%%%%%%%%%%%%%%%%%%%%%%%%%%%%%%%%%

\section{QNMs from massless test particles and photons in the eikonal approximation}

Motion of massless test particles and photons in the black hole geometry are described by the null geodesics. In the eikonal limit, i.e. when $ l >> 1$, the QNM frequencies of a black hole can be determined by the parameters of null geodesics \cite{cardoso1} \cite{mas1}. More specifically, the angular velocity at the unstable null geodesic, $\Omega_a$ and the Lyapunov exponent,  $\lambda_{Ly}$ are related to the QNM frequencies as follows:

\begin{equation} \label{nullfre}
\omega_{QNM} = \Omega_a l - i ( n + \frac{1}{2} ) | \lambda_{Ly}|,
\end{equation}
Here $\lambda_{Ly}$ in a dynamical system is a quantity that measure the average rate at which the nearby trajectories converge or diverge in the phase space. If a system has a positive Lyapunov exponent, it means that the nearby trajectories are diverging. Quantitative expressions for $\Omega_a$ and $\lambda_{Ly}$ will be defined at a later stage in this section. There are several papers which have employed this method for computing QNM frequencies \cite{dolan1} \cite{hod1} \cite{morgan} \cite{nora5}.

In the BIdS black hole space-time, for some sets of values of the parameters, the effective potential experienced by a massless test particle may present a maxima and therefore the possibility for the existence of unstable geodesics. The equation for the null geodesics in the black hole background is given by,
\begin{equation}
\dot{r}^2+V_{\rm null}=E^2, 
\end{equation}
where,
\be
\quad V_{\rm null}= \left({\frac{L^2}{r^2}}\right)f(r),
\ee
with $L$ denoting the angular momentum of the test particle and $f(r)$ is the metric function in eq. (\ref{metric}).
The radius at which the effective potential is maximum  is found using the condition that the first derivative of $V$ is zero; this leads to the equation,
\begin{equation} \label{null}
2f(r)-r f'(r)=0.
\end{equation}
The solution to eq$\refb{null}$  is denoted as $r_{\rm null}$. Eq$\refb{null}$ for the  BIdS metric is simplified to be,
\begin{equation}
r-3M+ Q^2 \sqrt{\frac{b}{Q}} F \left[{\arccos \left({\frac{r^2-Q/b}{r^2+Q/b}} \right), \frac{1}{\sqrt{2}}} \right]=0.
\label{r_null_massless}
\end{equation}

In the presence of nonlinear electromagnetic field, the behavior of photons and massless test particles are not the same.  Photons do not follow the null geodesics of the black hole metric. Instead, they follow the null geodesics  of an effective metric \cite{nov}  \cite{dud},  and unstable null geodesics of the corresponding modified effective potential  are not the same as the ones for the BIdS metric with eqs. (\ref{metric}) and (5). The effective metric in nonlinear electrodynamics, which determines the light ray trajectories, is given by
\begin{equation}
ds_{\rm eff}^2=\frac{G_m}{G_e} [-f(r) dt^2+f(r)^{-1} dr^2]+r^2 d\Omega^2  ,
\label{eff_metric}
\end{equation}
where the factors ${G_e},{G_m}$, correspond to the electric and magnetic charge modification introduced by the nonlinear field; they are given by
\begin{equation}
G_e=1- 4 L_{FF}\frac{Q_m^2}{L_F^3 r^4} , \quad G_m=1+ 4 L_{FF}\frac{Q_m^2}{L_F r^4},
\end{equation}
where $L(F)$ is the nonlinear electromagnetic Lagrangian and $L_F, L_{FF}$ are  its first and second derivatives with respect to $F$; here $F$ is the the electromagnetic invariant, $F=2(B^2-E^2)$. For the BIdS metric, with only electric charge, those factors are
\begin{equation}
G_m=1, \quad G_e=1 + \frac{Q^2}{b^2r^4}.
\end{equation}
Accordingly,  the effective potential for light rays (photons) in the nonlinear electromagnetic space-time is given by
\begin{equation}
V_{\rm photons}=\frac{G_e}{G_m}  \left({\frac{L^2}{r^2}}\right)f(r).
\end{equation}
Hence the equation for the radius corresponding to the maximum value of the potential, $r_{\rm null}$ will be modified as
\begin{equation}
\left({\frac{G_e}{G_m} }\right)\left({\frac{f'}{f}-\frac{2}{r}}\right)-\left({\frac{G_e}{G_m} }\right)'=0,
\end{equation}
which will lead to,
\begin{eqnarray}
&&3(r^5-3Mr^4-\frac{Q^2}{b^2} r+M\frac{Q^2}{b^2})+2 \Lambda \frac{Q^2}{b^2} r^3-4Q^2r[r^2- \sqrt{r^4+{Q^2}/{b^2}}] \nonumber\\
&&+Q^2(3r^4-\frac{Q^2}{b^2}) \sqrt{\frac{b}{Q}} F \left[{\arccos \left({\frac{r^2-Q/b}{r^2+Q/b}} \right), \frac{1}{\sqrt{2}}} \right]=0.
\label{r_null_ph}
\end{eqnarray}

\begin{figure} [H]
\begin{center}
\includegraphics{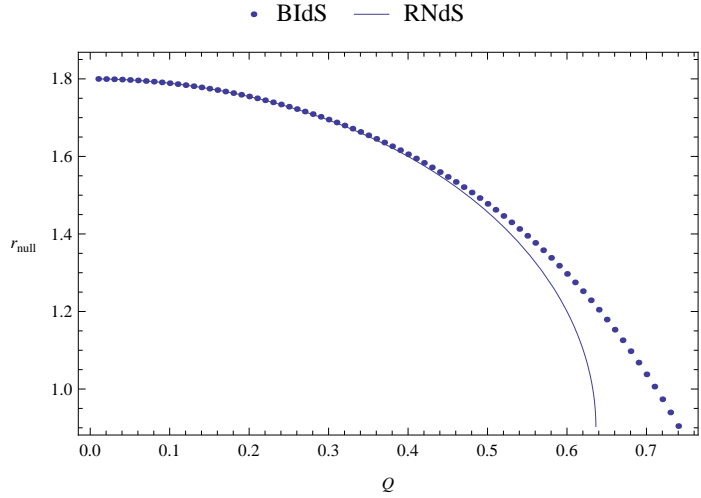}
\caption{The figure shows  $r_{ \rm null}$ vs $Q$ for massless particles. Here, $\Lambda = 0.1, b= 0.3$ and $M = 0.6$}
\label{Fig1}
 \end{center}
\end{figure}

\begin{figure} [H]
\begin{center}
\includegraphics{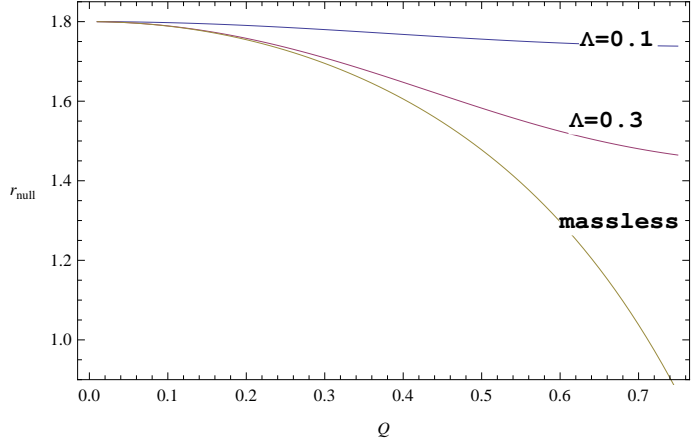}
\caption{The figure shows  $r_{\rm null}$ vs $Q$. The lower one is for massless particles and the upper two are for photons. Here, $M = 0.6, b=0.3$}
\label{Fig2}
 \end{center}
\end{figure}

\begin{figure} [H]
\begin{center}
\includegraphics{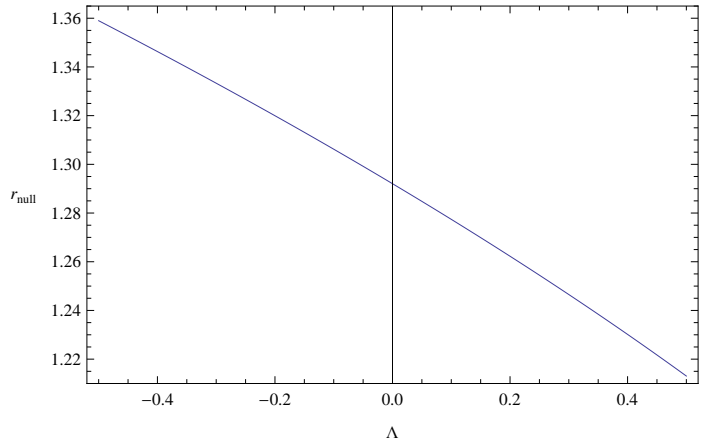}
\caption{The figure shows  $r_{\rm null}$ vs $\Lambda$. Here, $ M = 0.6, Q =0.6, b =1.4$ and $ -0.5 < \Lambda < 0.5$.}
\label{Fig3}
 \end{center}
\end{figure}

In Fig.$\refb{Fig1}$  $r_{\rm null}$ as a function of the charge $Q$ for massless particles is displayed. The maxima of the effective potentials for the RNdS and BIdS are compared as a function of $Q$. As $Q$ grows  $r_{\rm null}$ decreases.  The decreasing of  $r_{\rm null}$  indicates that the photosphere is closer to the horizon $r_h$.  Then for the same charge the photosphere is closer to the RNdS horizon than to the BIdS one. In Fig.$\refb{Fig2}$   $r_{\rm null}$ as a function of the charge $Q$ is shown for  BIdS. Comparison is done between  $r_{\rm null}$  for massless test particles and photons, as a function of $Q$  for different values of $\Lambda$. The two upper curves are for photons while the lower one is   for massless test particles. Photons are in fact affected when $\Lambda$ varies, while the maxima of the effective potential does not change with $\Lambda$  for massless test particles. This effect can be noticed by comparing eq.(\ref{r_null_ph}) for photons with the eq.(\ref{r_null_massless}) for massless particles, since $\Lambda$ does not appear in the latter.  As $Q$ grows  $r_{\rm null}$ decreases. In Fig.$\refb{Fig3}$ the behavior of  $r_{\rm null}$ for photons as a function of $\Lambda$ is shown. This is the plot of the root of eq.$\refb{r_null_ph}$.  In this case negative values of $\Lambda$ are also included for the sake of completeness. 
The values of the rest of parameters are fixed as $M=0.6, Q=0.6, b=1.4$, and $ -0.5 < \Lambda <0.5$.

%%%%%%%Lypunov ?%%%%%%%%%%

\begin{figure} [H]
\begin{center}
\includegraphics{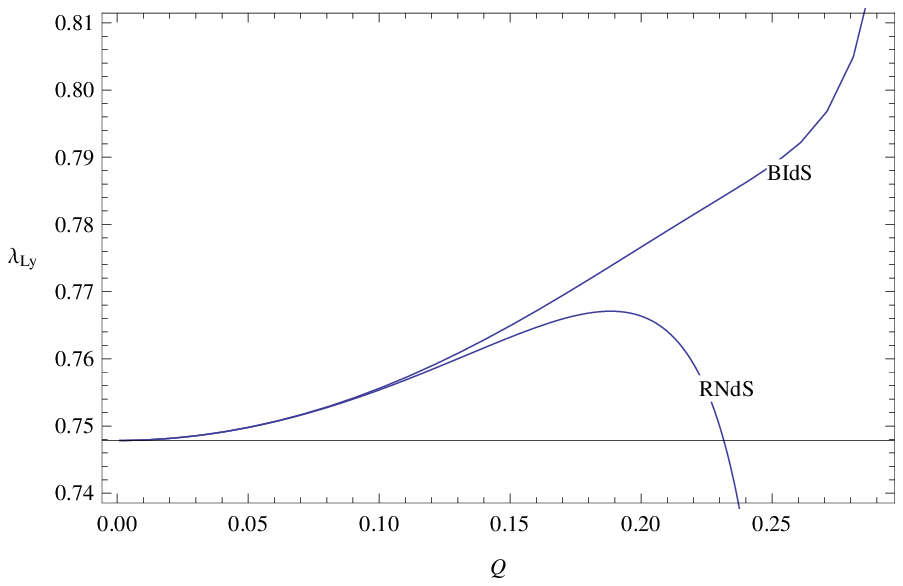}
\caption{The figure shows  $\lambda_{Ly}$ vs $Q$ for massless particles for BIds and RNdS black holes. Here $M =0.25, \Lambda =0.1, b =1$}
\label{Fig4}
 \end{center}
\end{figure}

\begin{figure} [H]
\begin{center}
\includegraphics{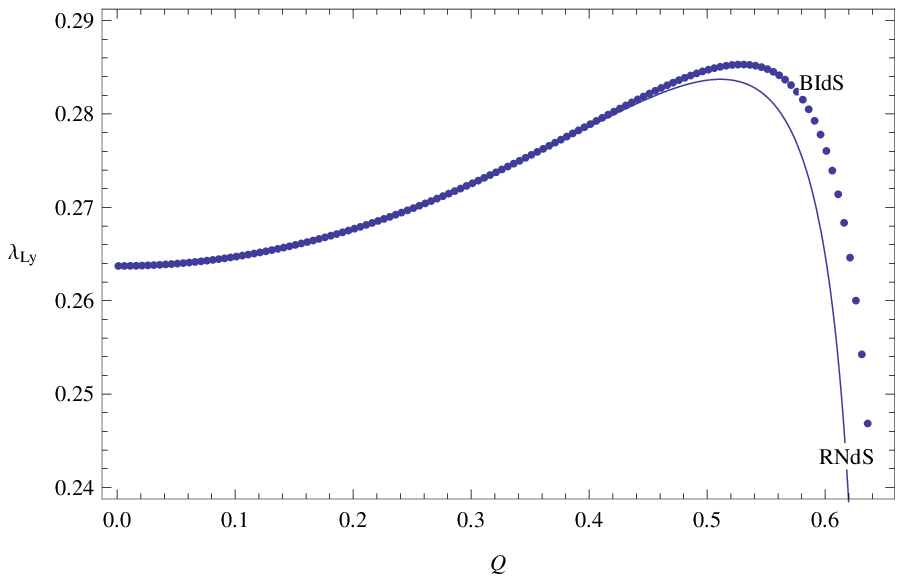}
\caption{The figure shows $\lambda_{Ly}$ vs $Q$ for massless particles for BIds and RNdS black holes. Here $M =0.6, \Lambda =0.1, b =1$ }
\label{Fig5}
 \end{center}
\end{figure}

\begin{figure} [H]
\begin{center}
\includegraphics{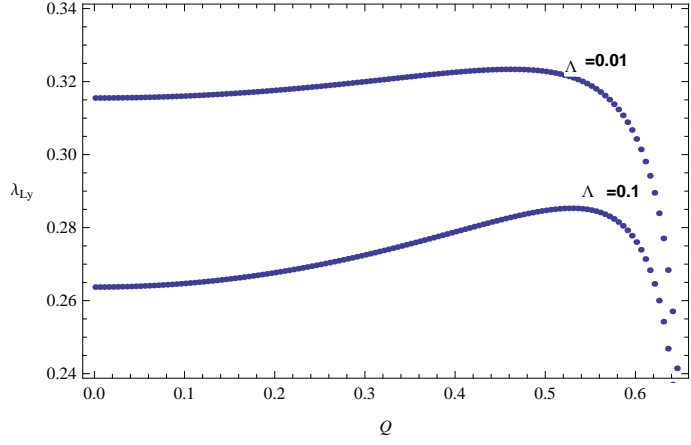}
\caption{The figure shows  $\lambda_{Ly}$ vs $Q$ for massless particles for the BIdS black hole for different values of $\Lambda$. Here $M =0.6, b =1$}
\label{Fig6}
 \end{center}
\end{figure}

\begin{figure} [H]
\begin{center}
\includegraphics{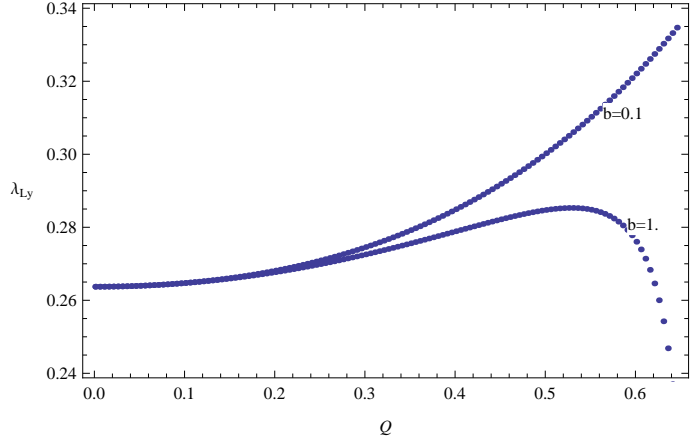}
\caption{The figure shows  $\lambda_{Ly}$ vs $Q$ massless particles for the BIdS black hole for two values of the BI parameter  $ b =0.1, 1$. Here $M =0.6, \Lambda=0.1$ }
\label{Fig7}
 \end{center}
\end{figure}

In the eikonal approximation the Lyapunov exponent for massless test particles is given by  

\begin{equation}
\lambda_{\rm Ly}= \sqrt{\frac{-V_{\rm null}'' (r_{\rm null})r_{\rm null}^2f(r_{\rm null})}{2L^2}}.
\label{Lya_massless}
\end{equation}
When  $V_{\rm null}'' $ for the BIdS metric  is substituted, the Lyapunov exponent is given by

\begin{equation}
\lambda_{\rm Ly}^2=\frac{f(r)}{r^2} \left({3-\frac{12M}{r}+ \frac{2Q^2}{\sqrt{r^4+ Q^2/b^2}}+ \frac{4Q^2}{r}\sqrt{\frac{b}{Q}} F \left[{\arccos \left({\frac{r^2-Q/b}{r^2+Q/b}} \right), \frac{1}{\sqrt{2}}} \right]}\right)
\label{V''_massless}
\end{equation}
The above  expression evaluated at $r_{\rm null}$.
In Figs.$\refb{Fig4}$,$\refb{Fig5}$, $\refb{Fig6}$ and $\refb{Fig7}$ the Lyapunov exponent is displayed for various values of parameters. In Fig.$\refb{Fig4}$ the Lyapunov exponents of BIdS and RNdS are compared for massless particles. For $Q=0$  both curves start at the same point, that is the Lyapunov exponent of the Schwarzschild-de Sitter black hole for the same mass. For this case, the values of the parameters are,  $M=0.25, \Lambda=0.1, b=1.$ The allowed range of $Q$ is shorter for RNdS than for BIdS: $0<Q<0.3$. Fig.$\refb{Fig5}$ is similar to Fig.$\refb{Fig4}$  but for a different value of the mass, $M=0.6$. As the mass grows, the behavior of the Lyapunov exponent of BIdS approaches the one of RNdS. 

For the light trajectories calculated from the effective metric, the Lyapunov exponent is 
\begin{equation}
\lambda_{\rm Ly}^2=\frac{f r^2}{2} \left[{ \frac{f}{r^2} \frac{G_m}{G_e}\left({\frac{G_e}{G_m} }\right)''- \left({\frac{f}{r^2} }\right)'' }\right]
\label{Lya_ph}
\end{equation}
here, $\lambda_{Ly}$ for  the BIdS black hole   is given by
\begin{eqnarray}
\lambda_{\rm Ly}^2 &= & \frac{f}{r^3 ({Q^2}/{b^2}+r^4)} \left[{- \frac{10}{3}\frac{Q^2}{b^2} \Lambda r^3+(8\frac{Q^2}{b^2}-12r^4) \left(\frac{Q^2}{3} \sqrt{\frac{b}{Q}}F(r)-M\right)} \right. \nonumber\\
&&  \left.    {+\frac{Q^2r}{3}[20r^2-26 \sqrt{{Q^2}/{b^2}+r^4}]+7\frac{Q^2}{b^2}r-3r^5}\right],
\label{Lya_factor}
\end{eqnarray}
with $F(r)$ denoting the elliptic function eq.(\ref{elliptic_hypergeom}).
 
In Fig.$\refb{Fig6}$ the Lyapunov exponent versus $Q$ is shown for two different values of $\Lambda$ for massless particles. As the charge grows the difference diminishes. In Fig.$\refb{Fig7}$  the Lyapunov exponent versus $Q$ for massless particles  is shown for two different values of $b$.   As the BI parameter $b$ grows the BIdS approaches the RNdS behavior.
So far we found agreement between the QNM at the  high-frequency limit and the ones obtained using the WKB method, that tells us that the frequency of the perturbation does not influence the black hole response. In summary, the black hole is stable for large $Q$ and small $\Lambda$. 

\begin{figure} [H]
\begin{center}
\includegraphics{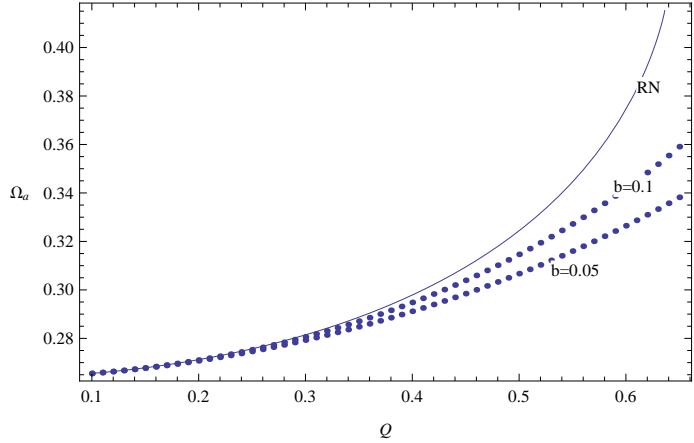}
\caption{The figure shows  $\Omega_a$ vs $Q$ for the  massless test particles. Here $ M =0.6, \Lambda =0.1$}
\label{Fig8}
 \end{center}
\end{figure}

\begin{figure} [H]
\begin{center}
\includegraphics{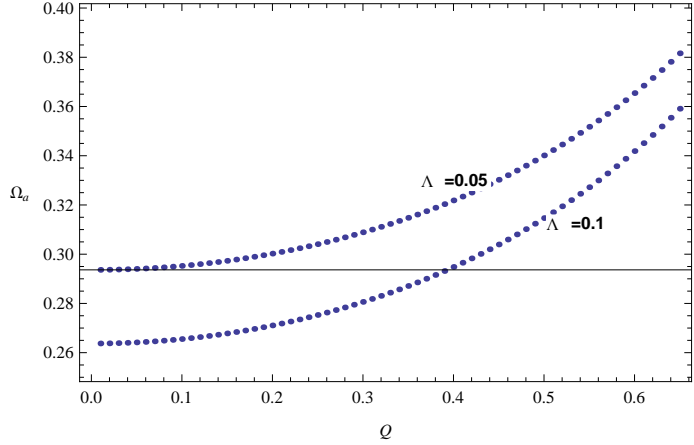}
\caption{The figure shows  $\Omega_a$ vs $Q$  for massless particles. Here $ M= 0.6, b =0.1$.}
\label{Fig9}
 \end{center}
\end{figure}

 In Figs.$\refb{Fig8}$ and Fig$\refb{Fig9}$  the real part of the QNM are displayed. They  are proportional to $\Omega_a$. The plots correspond to the following equations, for the massless test particles, 

\begin{equation}
\Omega_a^2=  \frac{f(r)}{r^2}
\label{Om_massless}
\end{equation}
evaluated at the roots of eq. (\ref{r_null_massless});
while for photons 
\begin{equation}
\Omega_a^2= \frac{G_m}{G_e} \frac{f(r)}{r^2}
\label{Om_ph}
\end{equation}
evaluated at the roots of eq.(\ref{r_null_ph}).
In Fig.$\refb{Fig8}$,  $\Omega_a$ of BIdS is plotted vs. $Q$ for two values of the BI parameter, $b=0.05$ and $b=0.1$ and compared with RNdS. As $b$ grows  $\Omega_a$ approaches the behavior of RNdS.
Nonlinear electromagnetism is to suppresses  $\Omega_a$.   In Fig.$\refb{Fig9}$  $\Omega_a$ versus $Q$ is shown for massless particles.  Comparison is done for two different values of $\Lambda$. As $\Lambda$ grows the value of  $\Omega_a$ decreases.

\begin{figure} [H]
\begin{center}
\includegraphics{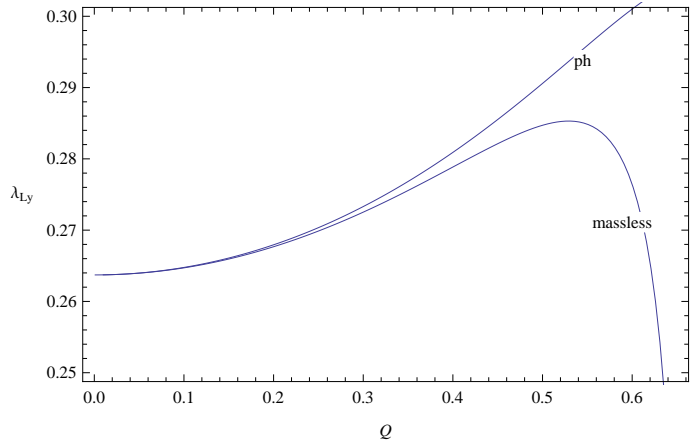}
\caption{The figure shows  $\lambda_{Ly}$ vs $Q$ for photons and massless test particles. $M =0.6, \Lambda =0.1, b=1$}
\label{Fig10}
 \end{center}
\end{figure}

\begin{figure} [H]
\begin{center}
\includegraphics{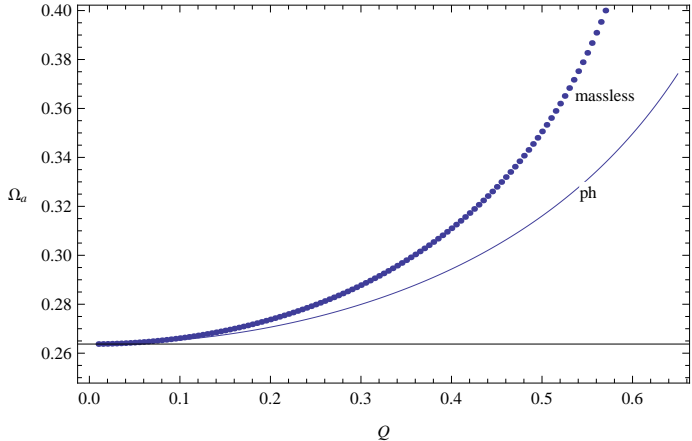}
\caption{The figure shows  $\Omega_{a}$ vs $Q$ for photons and massless particles. $M =0.6, \Lambda =0.1, b=1$}
\label{Fig11}
 \end{center}
\end{figure}

In Fig.$\refb{Fig10}$ the Lyapunov exponent corresponding to massless test particles and photons is compared. In Fig.$\refb{Fig11}$   $\Omega_a$  for massless test particles  is compared with the one of  light rays or photons.

%%%%%%%%%%%%%%Cross sections%%%%%%%%%%%

\section{ Absorption cross sections at low and high energy frequency limits}

Another important aspect to study when a black hole is perturbed by a test field is how much of the impinging field is absorbed by the black hole. 
Accretion rates and the growth of the mass of a black hole are related to the  magnitude of the absorption cross section. 
In this section, we will calculate the absorption cross section, in the low and high-frequency limits, for both the scalar and electromagnetic field via  null geodesics.

The low-frequency  absorption cross section for minimally-coupled scalar massless field of a static black hole has been computed to be equal to the horizon area of the black hole  by Das et.al. \cite{das}

\begin{figure} [H]
\begin{center}
\includegraphics{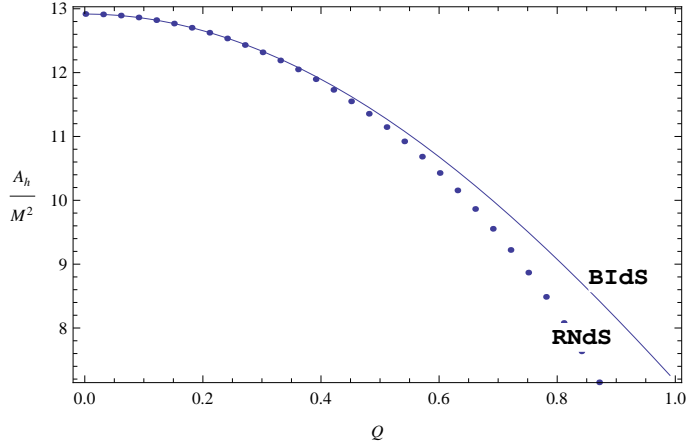}
\caption{The figure shows  the area of the event horizon ${A_h}/{M^2}$ vs $Q$. $M=1, \Lambda=0.01$ and $b=0.2$}
\label{Fig7b}
 \end{center}
\end{figure}

 In Fig$\refb{Fig7b}$  the area of the event horizon, $A= \pi r_h^2$ has been plotted for both the RNdS and the BIdS black holes for comparison, with  $M=1, \Lambda=0.01$ and $b=0.2$. The BIdS area turned out to be bigger than the one for RNdS. Hence, the absorption cross section is larger for the BIdS black hole. In this case a distinction is not made between photons and massless test particles, since the horizon is the same for both particles.

In the high-frequency limit the absorption cross section can be considered as the classical capture cross section generated by the null geodesics in the case of massless  scalar waves. In this limit the absorption cross section is also called geometric cross section, $\sigma_{\rm geo}$, and   it is given by

\begin{equation}
\sigma_{\rm geo(scalar)}= \pi \beta_{\rm null}^2=  \frac{ \pi r_{\rm null}^2}{f(r_{\rm null})},
\end{equation}
Here, $r_{\rm null}$ is the radius of the unstable circular orbit obtained from $\frac{dV_{\rm null}}{dr}=0$, and  $\beta$ is the impact parameter given by $\beta = \frac{L}{E}$ where $L$ is the angular momentum and $E$ is the energy of the test particle at infinity. $\beta_{\rm null}$ corresponds to the value for the unstable circular orbit. $V_{\rm null}$ and $\beta_{\rm null}$ are related by,
\begin{equation}
V_{\rm null}=\frac{E_{\rm null}^2}{L_{\rm null}^2}=\frac{1}{\beta_{\rm null}^2}.
\end{equation}
 For photons, to calculate $\sigma_{geo}$, one has to consider the effective metric as was done in section(5). Hence the corresponding geometric cross section for the photon is given by,
 \begin{equation}
\sigma_{\rm geo(photons)}= \frac{ \pi r_{\rm null}^2}{f(r_{\rm null})G_e(r_{\rm null})},
\end{equation}

\begin{figure} [H]
\begin{center}
\includegraphics{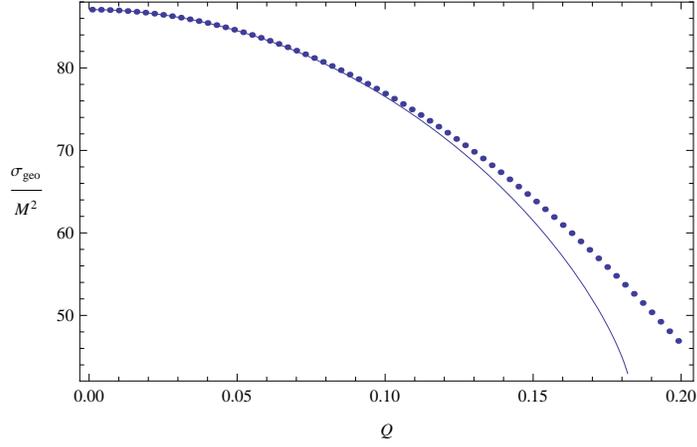}
\caption{The figure shows  $\frac{\sigma_{geo}}{M^2}$ vs $Q$ for the massless scalar for BIdS and RNdS black holes. The continuos curve is for RNdS and the doted one is for BIdS black hole. Here $M=0.01715, \Lambda=0.0973, b=0.5$ }
\label{Fig1b}
 \end{center}
\end{figure}

\begin{figure} [H]
\begin{center}
\includegraphics{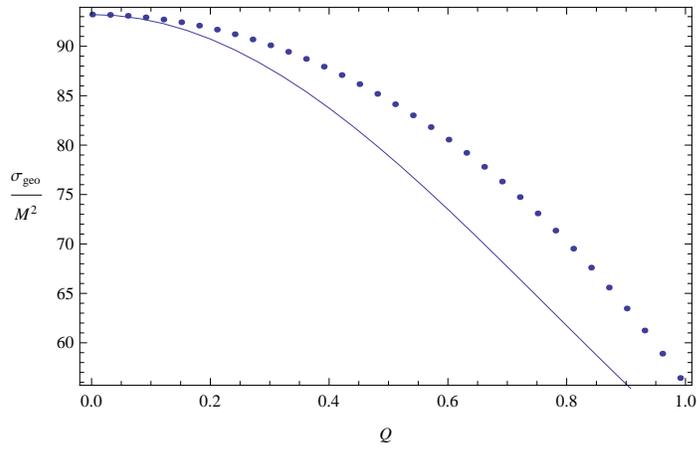}
\caption{The figure shows  ${\sigma_{geo}}/{M^2}$ vs $Q$ for the massless scalar field and the photons for the BIdS black hole. The continuos curve is for the photons and dotted is for the scalar field. Here $M=1, \Lambda=0.01$,  and $b=0.2$ }
\label{Fig4b}
 \end{center}
\end{figure}

\begin{figure} [H]
\begin{center}
\includegraphics{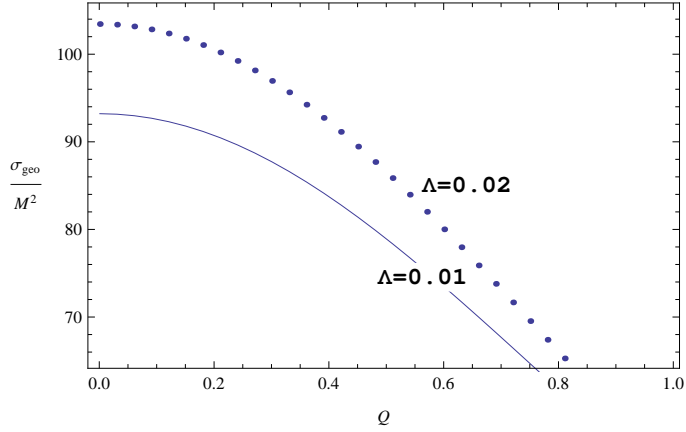}
\caption{The figure shows  ${\sigma_{geo}}/{M^2}$ vs $Q$ for  photons for two values of $\Lambda$. Here $M=1, b=0.2$ }
\label{Fig5b}
 \end{center}
\end{figure}

In Fig$\refb{Fig1b}$ the geometric cross section for massless test particles of  RNdS and BIdS  is illustrated  as a function of the electric charge $Q$. The continuous curve is for the RNdS black hole while the dotted is for BIdS; $\sigma_{\rm geo}$ is smaller for RNdS than the one of BIdS.  In Fig$\refb{Fig4b}$ the geometric cross section for BIdS comparing the one for  massless test particles with the one corresponding to photons
 is illustrated  as a function of the electric charge $Q$. The continuous curve corresponds to photons while the dotted one is for massless test particles. The absorption is higher for massless particles. In Fig$\refb{Fig5b}$  the geometric cross section from BIdS black hole for photons   is illustrated  as a function of the electric charge $Q$ and the comparison is done between two different values of $\Lambda$, $\Lambda=0.01$ and $\Lambda=0.02$, as the labels on each curve show. The geometric cross section increases with $\Lambda$.

Decanini et.al \cite{deca} showed that the oscillatory part of the absorption cross section in the eikonal limit can be written in terms of the parameters of the unstable null circular orbits, the Lyapunov exponent $\lambda_{\rm Ly}$ and the angular velocity $\Omega_a$ as follows:

\begin{equation}
\sigma_{\rm osc}=- \frac{4 \pi \lambda_{\rm Ly}}{\omega \Omega_a^2} e^{-\frac{\pi \lambda_{\rm Ly}}{ \Omega_a}} \sin\left({\frac{2 \pi \omega}{\Omega_a}}\right),
\end{equation}
where $\omega$ is the oscillating frequency of the scalar field.
Therefore the high-frequency limit of the absorption cross section is proportional to the sum of  $\sigma_{\rm osc}$ and $\sigma_{\rm geo}$,

\begin{equation}
\sigma_{\rm abs}^{\rm hf}=\frac{\pi r_{\rm null}^2}{f(r_{\rm null})} \left(1- \frac{4 \lambda_{\rm Ly}}{\omega} e^ {- \frac{\pi \lambda_{\rm Ly}}{ \Omega_a}} \sin\left({\frac{2 \pi \omega}{\Omega_a}}\right) \right),
\end{equation}
The above is known as the sinc approximation in the literature.
\noi
Macedo and Crispino studied absorption cross sections for a massless scalar wave in Bardeen black hole \cite{cris}. They calculated absorption cross section via the sinc approximation as well as numerically solving the equations for each $\omega$. For Bardeen black hole they showed that the sinc approximation agree remarkably well not only at high frequency limit, but also at the intermediate frequencies.

\begin{figure} [H]
\begin{center}
\includegraphics{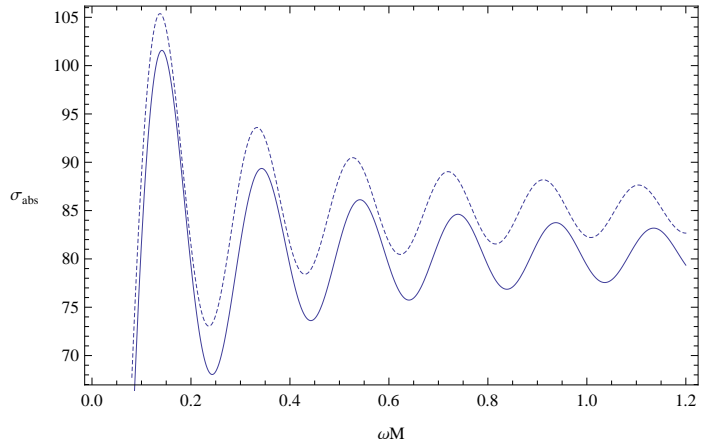}
\caption{The figure shows  $\sigma_{abs}$ vs $\omega M$ for the massless scalar field for RNdS and BIdS. The continuous curve is for the RNdS black hole while the dotted one  is for BIdS. Here,  $M=1, \Lambda=0.01, Q=0.6$ and $b=0.01$}
\label{Fig2b}
 \end{center}
\end{figure}

\begin{figure} [H]
\begin{center}
\includegraphics{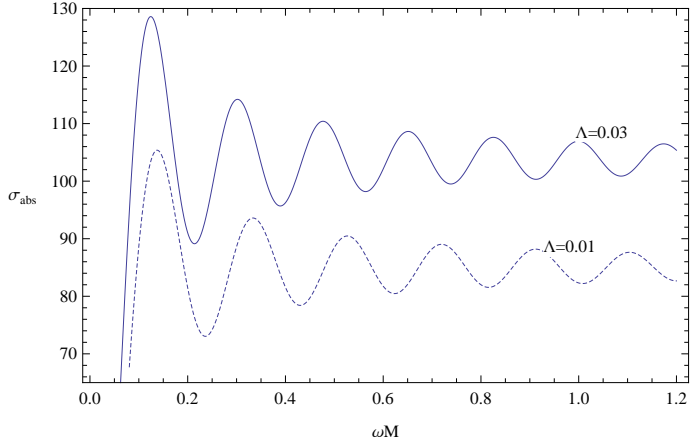}
\caption{The figure shows  $\sigma_{abs}$ vs $\omega M$ for the massless scalar field for the BIdS.  Here,  $M=1, b=0.01, Q=0.6.$}
\label{Fig3b}
 \end{center}
\end{figure}

\begin{figure} [H]
\begin{center}
\includegraphics{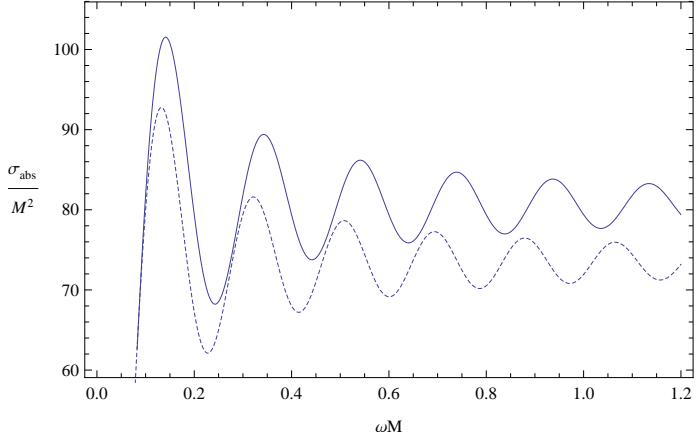}
\caption{The figure shows  $\sigma_{abs}$ vs $\omega M$  for the BIdS.  Here. the massless test particles is given in continuous curve with the corresponding one for photons given by the  dotted curve. Here,  $M=1, \Lambda=0.01, Q=0.6, b =0.2$}
\label{Fig6b}
 \end{center}
\end{figure}

In the Fig$\refb{Fig2b}$ the absorption cross section in the sinc approximation is presented  for massless test particles for  RNdS and BIdS, as a function of the frequency $\omega M$. The BIdS absorption cross section is greater than the RNdS. In the Fig$\refb{Fig3b}$  the absorption cross section in the sinc approximation from massless test particles impinging on the BIdS black holes, as a function of the frequency $\omega M$, is illustrated for two values of $\Lambda$, $\Lambda=0.01$ and $\Lambda=0.02$.   As $\Lambda$ increases $\sigma_{\rm abs}$ also increases. In the Fig$\refb{Fig6b}$ the absorption cross section for BIdS  as a function of 
$\omega M$  is illustrated,  comparing  the one for  massless test particles  and the photons. It can be seen that  $\sigma_{\rm abs}$ for photons is suppressed as compared to massless particles, i.e. the latter are more easily captured by the BIdS black hole (Figs. \ref{Fig4b} and \ref{Fig6b}).

%%%%%%%%%%%%%%%%%%%%%

\section{Conclusions}

In this paper, we have studied Born-Infeld black hole with a positive cosmological constant (BIdS). These black holes have interesting characteristics so that both time-like and space-like singularities exists depending on the parameters chosen. The main purpose of this paper has been to study the QNM's for the massless and photons. We also have presented absorption cross sections.

First massless scalar fields perturbation is analyzed with the WKB approximation. QNM frequencies were calculated for all parameters, $M, Q, \Lambda, l$ and $b$. When $Q$ is increased, $\omega_r, \omega_i$ and the quality factor increases; BIdS black hole is stable for large charge and it is a better oscillator for large charge.  When $\Lambda$ increases, $\omega_r, \omega_i$ and the quality factor decreases; the black hole is more stable for smaller $\Lambda$. When the non-linear parameter $b$ is increased, $\omega_r$ increases to reach a stable value; $\omega_i$ increases to a maximum and decreases for large $b$. Hence there is value of $b$ that the black hole reach it's maximum stability. The quality factor also increases with $b$. Hence RNdS black hole is a better oscillator compared to the BIdS black hole.

When $\omega$ is analyzed with respect to $l$, we observed that $\omega_r$ increases linearly with respect to $l$; $\omega_i$ decreases and reach a stable value for larger $l$. Hence the black hole is more stable for smaller values of $l$. We also computed $\omega$  with respect to $n$, the number of harmonics. When $n$ increases, $\omega_r$ decreases and $\omega_i$ increases; the fundamental frequency takes the longest time to decay and the oscillating frequency is higher for smaller harmonics.

All the $WKB$ calculations were done for $ l >0$ values. We observed that for $l =0$, the potential has a negative region and a local minima  in between the horizons; one cannot apply WKB method to compute QNM frequencies of $ l =0$ case. It would be interesting to apply another method to study the QNM frequencies and see if the BIdS black hole is stable for $l =0$ case.

In the second half of the paper null geodesic  of the black hole were employed to study the QNM frequencies  and the absorption cross sections at the eikonal limit.  The real part of $\omega$ is proportional to  the angular velocity of the null geodesics, $\Omega_a$. The imaginary part of $\omega$ is proportional to the Lyapunov coefficient, $\lambda_{Ly}$.

First, QNM frequencies for both massless and photons were calculated via the angular velocity and the Lyapunov coefficient of the null geodesics. In black holes with non-linear electrodynamics, the photons follow null geodesics of an effective geometry. The massless particles follow the usual null geodesics of the black hole geometry.  

When $Q$ is increased QNM's for the BIdS increases and QNM's for the RNdS increases to a maximum  and then decreases. When observed $\omega_i$ with respect to $Q$, higher value of $\Lambda$ has lower $\omega_i$ similar to what was obtained with the WKB analysis. For massless particles, $\omega_r$ increases with $Q$ and the value is higher for larger $b$ (for the same $Q$). RNdS black hole has a larger $\omega_r$ than the BIdS black hole. When plotted for BIdS black hole alone, $\omega_r$ is smaller for larger value of $\Lambda$.

When $\lambda_{Ly}$ is plotted, the photons have a higher value than the massless particles. Hence the black hole is more stable for electromagnetic perturbations. On the other hand $\omega_r$ is larger for the massless particles when compared with the photons.

The absorption cross sections were studied for both BIdS and RNdS black holes. At low frequencies, the absorption cross section for massless particles  is higher for the BIdS; BIdS black hole is a better absorber of the massless scalar waves.

At higher frequencies, the absorption cross section is given by the geometric cross section. Here BIdS black hole has a larger $\sigma_{geo}$ compared to RNdS for massless particles. For the BIdS black hole, $\sigma_{geo}$ is larger for the massless scalar field than for the photons. Computations are done for the BIdS by varying $\Lambda$. When $\Lambda$ is larger, $\sigma_{geo}$ is larger.

The high frequency limit of the absorption cross section was also calculated by the sink approximation given by $\sigma_{abs}$. When $\sigma_{abs}$ vs $ \omega M$ is plotted for the massless particles, $\sigma_{abs}$ is higher for the BIdS black hole. Also, $\sigma_{abs}$ is larger for higher $\Lambda$ for the BIdS black hole. Hence the BIdS absorbed more for larger cosmological constant. When  $\sigma_{abs}$ is compared for the photons and the massless particles,  the massless particles has higher absorption.

%%%%%%%%%%%%%%%%%%%%%%%%%%%%%%%
\vspace{0.5 cm}

%%%%%%%%%%%%%%%%%%%%%

{\bf Acknowledgments:}  SF  wish to thank R. A. Konoplya for providing the  {\it Mathematica}  file for  the WKB approximation.

%%%%%%%%%%%%%%%%%%%%%%

\end{document}